\documentclass[reprint,showpacs,amsmath,amssymb,twocolumn,letterpaper,aps,prb]{revtex4-1}
\usepackage{textcomp}
\usepackage{graphicx}
\usepackage{amsmath}
\usepackage{amssymb}
\usepackage{times}

\newcommand{\ua}
{UCo$_{1-x}$Fe$_x$Ge}

\begin{document}
\title{Ferromagnetic Quantum Critical Point in UCo$_{1-x}$Fe$_x$Ge}

\author{K. Huang$^1$}
\author{J. J. Hamlin$^1$}
\author{R. E. Baumbach$^1$}\altaffiliation[Present Address: ]{Condensed Matter and Magnet Science group, Los Alamos National Laboratory, Los Alamos, New Mexico 87545}
\author{M. Janoschek$^1$}\altaffiliation[Present Address: ]{Condensed Matter and Magnet Science group, Los Alamos National Laboratory, Los Alamos, New Mexico 87545}
\author{N. Kanchanavatee$^1$}
\author{D. A. Zocco$^1$}
\author{F. Ronning$^2$}
\author{M. B. Maple$^1$}
\email[Corresponding Author: ]{mbmaple@ucsd.edu}

\affiliation{$^1$Department of Physics, University of California, San Diego, La Jolla, California 92093}
\affiliation{$^2$MPA-CMMS Los Alamos National Laboratory, Los Alamos, NM 87545\\}

\begin{abstract}
We have carried out a comprehensive study of the UCo$_{1-x}$Fe$_x$Ge series across the entire range of compositions $0 \leq x \leq 1$, and report the results of x-ray diffraction, magnetization, specific heat, and electrical resistivity measurements to uncover the $T-x$ phase diagram. Substitution of Fe into UCoGe initially results in an increase in the Curie temperature and a rapid destruction of the superconductivity.  Near $x=0.22$, the ferromagnetic transition is suppressed to zero temperature at an apparent ferromagnetic itinerant electron quantum critical point, where the temperature dependence of the electrical resistivity and specific heat in this region reveal non-Fermi liquid behavior.
\end{abstract}

\pacs{add pacs}

\maketitle
\section{Introduction}

The topic of quantum criticality continuously raises scientific interest because it is believed to be at the heart of the physics of emergent phenomena such as unconventional superconductivity, hidden order and the breakdown of the Fermi liquid model.\cite{Maple:10,stewart_2001} The phase transition from the disordered paramagnetic to ordered ferromagnetic state (FM) in zero magnetic field is a prototypical example of a critical point, i.e., a second-order phase transition. The inherently low Curie temperatures of metallic FMs makes them ideal systems to investigate magnetic quantum criticality, because the Curie temperature $T_C$ is often easily suppressed to zero temperature using an external tuning parameter such as pressure $P$ or chemical composition $x$, resulting in a FM quantum critical point (QCP).

However, detailed experimental work on archetypal FM metals such as MnSi,\cite{Pfleiderer:97} UGe$_2$\cite{Pfleiderer:02}, and ZrZn$_2$\cite{Uhlarz:04}, has demonstrated that the situation is more complex than this simple model would suggest. For these compounds, the FM phase is suppressed to zero temperature at a first-order transition; i.e., no quantum critical behavior is observed. This can be understood theoretically in various ways: e.g., (1) additional fermionic modes may couple to the FM critical fluctuations and are expected to generically drive the phase transition to first-order,\cite{Belitz:99} or (2) magnetoelastic coupling may cause a phase transition to become first order.~\cite{Mineev:2011} On the other hand, each of these systems exhibit additional complicating factors that make it difficult to derive the universal behavior of FM quantum phase transitions (QPTs); MnSi is a long-period helimagnet (therefore only locally FM) in which the thermal phase transition is weakly first-order,\cite{Janoschek:12} UGe$_2$ is a strongly uniaxial FM, and ZrZn$_2$ exhibits a marginal Fermi liquid ground state.\cite{Sutherland:12}

These difficulties are further underlined by URu$_{2-x}$Re$_x$Si$_2$, which is believed to exhibit a FM QCP.\cite{Butch:09} URu$_{2-x}$Re$_x$Si$_2$ has its own set of peculiarities, including unconventional critical scaling of the magnetization, non-Fermi-liquid behavior not only in vicinity of the QCP but also within the FM phase, and $\omega/T$ scaling of the magnetic fluctuations not expected for conventional QCP scenarios in FMs.\cite{Krishnamurthy:08} A further issue may be the disorder associated with tuning via chemical substitution, which, in principle, may smear out the phase transition. We note that at least some of the unconventional behavior of URu$_{2-x}$Re$_x$Si$_2$ may be understood within a theory that describes a FM Kondo lattice.\cite{Yamamoto:10} However, it is still unclear whether quantum phase transitions of FM metals are universally first-order. Moreover, as pointed out by Pfleiderer, both first- and second-order QPTs may lead to interesting new phenomena in metallic systems.\cite{Pfleiderer:05}

The interest in the nature of FM QPTs has been further intensified by the recent discovery of unconventional superconductivity (SC) in uranium-based compounds. The compounds UGe$_2$ (under pressure),\cite{BAUER01, ucoge:saxena} URhGe,\cite{ucoge:aoki} UIr (under pressure),\cite{akazawa_2004_1} and UCoGe\cite{ucoge:huy} exhibit microscopic coexistence of SC and itinerant electron FM that is apparently associated with the uranium ions. This has led to the suggestion that the SCing electrons in the uranium-based FM superconductors may pair in triplet states with parallel alignment of the electron spins, since such a configuration may not be subject to the same magnetic pair-breaking effects for a singlet SC (although orbital pairbreaking could still be present). Furthermore, it has been proposed that the SC in these compounds is mediated by critical fluctuations associated with a FM QCP.\cite{ucoge:huy,ucoge:nijs_1,Stock:11,Hattori:12} Naturally, this raises the question of how the proposed first-order nature of FM phase transitions influences the SC state, since in that case critical fluctuations will be absent. In addition, SC with triplet pairing is thought to be extremely sensitive to disorder,\cite{Fay:80} demonstrating the need for a model system in which disorder effects on FM SC may be studied systematically.

In the present manuscript, we expand on our recent results for the alloy series UCo$_{1-x}$Fe$_x$Ge which may shed some light on these questions.\cite{hamlin_2010} We report transport, magnetic, and thermal properties, which reveal an evolution from ferromagnetism in UCoGe towards Pauli paramagnetism in UFeGe.\cite{ucoge:canepa} In agreement with a recent report which was restricted to small iron dopant concentrations,\cite{pospisil_2009_1} we find that $T_C$ initially increases and the superconducting critical temperature $T_s$ is rapidly suppressed. At higher iron concentrations, $T_C$, the ordered magnetic moment $M_o$, and the magnetic entropy $S_{mag}(T_C)$ pass through a maximum and are then suppressed smoothly towards a quantum phase transition near $x_{cr}=0.22$. Non-Fermi-liquid (NFL) behavior is observed in the electrical resistivity and specific heat near $x_{cr}$. We also find that $T_C$ is suppressed with Fe substitution in a manner that is consistent with theoretical expectations for a three dimensional itinerant electron ferromagnet.\cite{hertz_1976,millis_1993} These findings point to the presence of a FM QCP at $x_{cr}$ where deviations from theoretical predictions in the critical low $T$ behaviors are likely due to disorder. Remarkably, we also find that samples with $x = 0$ and 0.2 are very similar, suggesting that apart from the absence of SC (above 50 mK, the lowest temperature to which the samples were measured) around $x_{cr}$, the FM dome is symmetric with respect to its magnetic properties. The substitutional system UCo$_{1-x}$Fe$_x$Ge thus presents a unique opportunity to investigate how disorder influences FM quantum critical fluctuations and their resulting emergent phenomena, namely superconductivity and/or non-Fermi-liquid behavior, by comparing the QCP situated near $x=0$ (accessible via pressure\cite{hassinger_2008_1,slooten_2009_1}) with the QCP at $x_{cr}$ that is more affected by disorder.

\section{Experimental Details}
Samples of \ua, with various concentrations spanning the range from $x = 0$ to 1, were prepared using an arc furnace in a zirconium gettered argon atmosphere on a water cooled copper hearth. The starting materials were cut from solid pieces of pure uranium (New Brunswick Lab, 3N), cobalt (Alfa Aesar, 99.9+$\%$), iron (Alfa Aesar, 99.99+$\%$) and germanium (Alfa Aesar Puratronic, 99.9999+$\%$).  After melting, each sample was flipped over and re-melted.  This process was repeated five times in order to ensure homogeneous mixing of the starting materials.  Each as-grown boule was cut in half using a diamond wheel saw, and half of the sample was sealed under vacuum in a quartz tube and annealed at 900$^{\circ}$C for 10 days. Following our initial studies of polycrystalline material, we produced a single crystal specimen with $x$ $=$ 0.22 using the Czochralski technique. In the following figures, we use stars to identify this sample.
\begin{figure}
  \begin{center}
    \includegraphics[width=1\columnwidth]{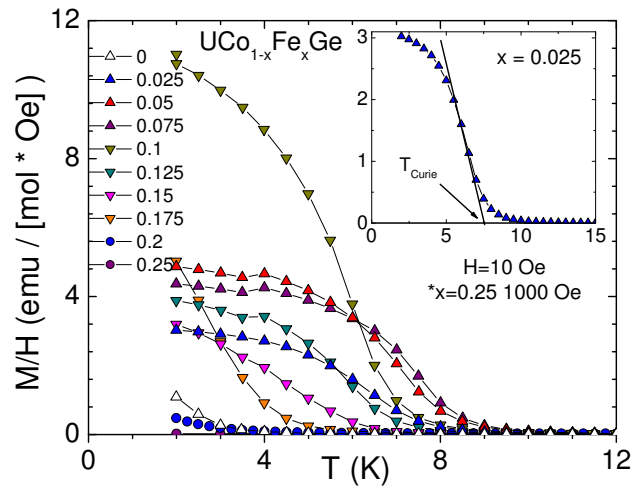}
  \end{center}
  \caption{Magnetization $M$ divided by applied field $H$ versus temperature $T$ for \ua\ with $x$ = 0 - 0.25. All scans were measured upon field cooling in 10 Oe (except for $x$ = 0.25 in 1000 Oe). The inset shows an example of the construction used to estimate $T_C$.}
  \label{fig:fig1}
\end{figure}

Powder diffraction patterns were obtained with a Bruker D8 diffractometer utilizing Cu K$_{\alpha}$ radiation and the results analyzed via Rietveld refinement using the GSAS+EXPGUI\cite{rietveld_1969_1,larson_2000_1} software package.  Magnetization data for temperatures 2-300 K and fields up to 7 tesla were collected with a Quantum Design Magnetic Properties Measurement System (MPMS) SQUID magnetometer.  Resistivity measurements were performed using a standard four wire method with a Linear Research LR-700 AC resistance bridge operating at $\sim 16$ Hz.  For the resistivity measurements, temperatures down to 1 K were achieved using a home-built $^4$He bath cryostat while data below 1 K were taken in an Oxford Kelvinox dilution refrigerator.  Several resistivity measurements were repeated with varying currents in order to ensure that spurious heating did not effect the data at the lowest temperatures.  Specific heat measurements were obtained from 1.8-50 K with a Quantum Design Dynacool Physical Properties Measurement System (PPMS) utilizing a standard adiabatic heat pulse technique. Specific heat measurements were obtained from 0.4-20 K using a Quantum Design Physical Property Measurement System with a He-3 option.

\section{Results}
While UCoGe exhibits the orthorhombic TiNiSi crystal structure, UFeGe undergoes a monoclinic distortion of the TiNiSi structure below $\sim 500$~$^{\circ}$C.\cite{ucoge:canepa} We found that the orthorhombic TiNiSi structure of UCoGe persisted up to $x < 70\%$ for UCo$_{1-x}$Fe$_x$Ge.~\cite{hamlin_2010}  The $x$ = 0.7, 0.8, and 0.9 samples showed significant impurity phases and their diffraction patterns could not be fitted well by the calculated patterns for orthorhombic TiNiSi, monoclinic UFeGe, or a mixture of these phases.  The as-grown doped samples ($x < 0.7$) showed only small concentrations of an unidentified impurity phase (the integrated intensity of the largest impurity peak is only $\lesssim 2 \%$ of the largest sample peak).  Upon annealing, the size of the impurity peaks grows significantly, indicating that annealing of doped \ua\ results in significant precipitation of impurity phases.  For pure UCoGe, annealed samples have previously been reported to possess significantly larger residual resistivity ratios than as-grown samples.\cite{huy_2009_3}  For the above reasons, we restrict our analysis to annealed pure UCoGe and as-grown doped \ua\ samples (with $x < 0.7$).
\begin{figure}
  \begin{center}
    \includegraphics[width=1\columnwidth]{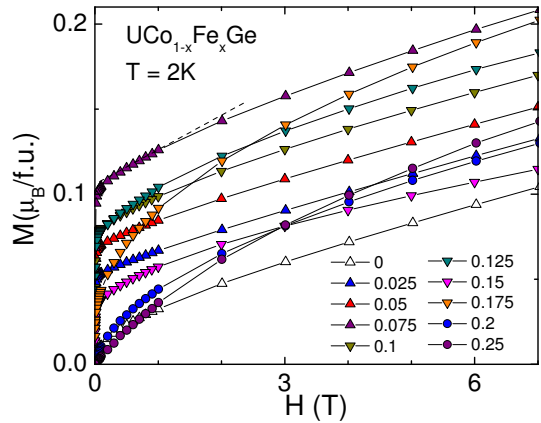}
  \end{center}
  \caption{Magnetization $M$ versus applied field $H$ at 2 K.  The dashed line shows an example of the construction used to estimate the ordered magnetic moment $M _o$.}
  \label{fig:fig2}
\end{figure}

In Figure~\ref{fig:fig1}, we plot the low temperature magnetization $M$ divided by magnetic field $H$ versus temperature $T$ data, from which the evolution of the Curie temperature $T_C$ is estimated by extrapolating $M$ below the ordering temperature to zero as shown in the inset.  As in previously reported measurements,\cite{ucoge:huy} the magnetization of pure UCoGe exhibits a weak upturn at low temperatures, consistent with FM ordering at $\sim 3$ K.  Figure~\ref{fig:fig7}(a) shows $T_C$ versus $x$, demonstrating that $T_C$ initially increases rapidly, passes through a dome with a maximum of $\sim 9$ K at $x \sim 0.075$, and then drops to zero between $x \sim 0.2-0.25$. We also find that the value of $M/H$ at $T$ $=$ 2 K passes through a dome as $x$ is varied with a maximum near $x$ $=$ 0.1, suggesting that Fe substitution initially drives the magnetism from being itinerant near $x$ $=$ 0 to more localized near $x$ $=$ 0.1, after which it again becomes itinerant near $x$ $=$ 0.2.

Further insight into the magnetic state is provided by the $M$ versus $H$ measurements presented in Figure~\ref{fig:fig2}.  These data were taken upon decreasing the field from 7 tesla, after cooling the samples to 2 K in zero field.  For concentrations $x \leq 0.2$, the data show clear FM characteristics, while for concentrations $x \geq 0.25$, the behavior is consistent with paramagnetism.  A rough estimate of the ordered magnetic moment $M _{o}$ is obtained by linearly extrapolating the high field data to zero field as shown for $x = 0.075$ by the dashed line construction in Figure~\ref{fig:fig2}.  The resulting values of the ordered moments are plotted versus $x$ in Figure~\ref{fig:fig7}(c), demonstrating that $M _{o}$ passes through a dome, peaked at $x = 0.075 - 0.1$, and goes smoothly to zero near $x = 0.2 - 0.25$.  The value of $M _{o}$ at $x = 0.075 - 0.1$ is more than an order of magnitude larger than in undoped UCoGe, indicating a significant strengthening of the magnetic state, while at $x = 0.2$, the ordered moment has dropped to nearly the same value as in undoped UCoGe, suggesting a return to highly itinerant weak ferromagnetism.
\begin{figure}
  \begin{center}
    \includegraphics[width=1\columnwidth]{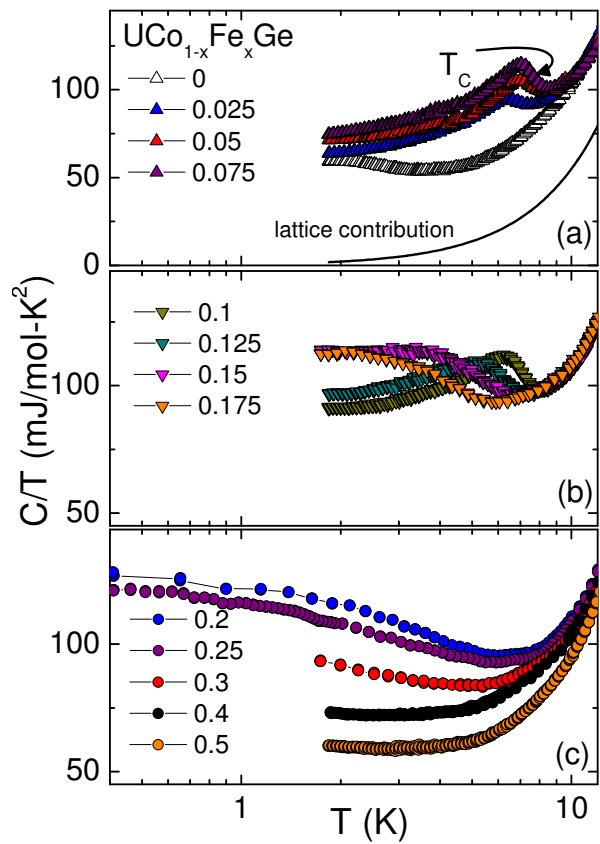}
  \end{center}
  \caption{(a) Specific heat $C$ divided by temperature $T$ versus $T$ on a log scale in zero magnetic field $H$ for concentrations 0 $\leq$ $x$ $\leq$ 0.075.  The arrow indicates the Curie temperature estimated for the $x=0.075$ sample. The solid line is the estimated lattice contribution to the data ($C_{lat}/T$ $=$ $\beta$$T^2$) for $T_C$ $\leq$ $T$ $\leq$ 15 K where $\beta$ $\approx$ 0.55 mJ/mol-K$^4$ from fits to the data. (b) $C/T$ versus $T$ on a log scale in zero magnetic field $H$ for concentrations 0.1 $\leq$ $x$ $\leq$ 0.175. (c) $C/T$ versus $T$ on a log scale in zero magnetic field $H$ for concentrations 0.2 $\leq$ $x$ $\leq$ 0.5.}
  \label{fig:fig3}
\end{figure}

Figure~\ref{fig:fig3} presents the specific heat $C$ divided by temperature $T$ versus $T$ on a log scale.  Pure UCoGe shows a weak anomaly with an onset near 3 K, and samples with $0.025 \leq x \leq 0.175$ show clear anomalies that are coincident with the onset of ferromagnetic order.  The ordering temperatures estimated from the inflection point at the anomaly are plotted versus $x$ in Figure~\ref{fig:fig7}(a) and are in agreement with results from magnetization. The data for $x$ = 0.2, 0.25, and 0.3 show nearly logarithmic low $T$ upturns, but saturate towards constant values at low $T$ indicating the presence of a nearby QCP which induces NFL behavior.  The value of $C/T$ at $T$ $=$ 1 K, which offers a measure of the effective charge carrier quasiparticle mass, is plotted versus $x$ in Figure~\ref{fig:fig7}(d). Here, we find that $C_{1K}/T$ reaches a maximum between $x$ = 0.2 and 0.25, consistent with the viewpoint that strong spin fluctuations are present near the region where the ferromagnetism is suppressed towards $T$ $=$ 0.

\begin{figure}
  \begin{center}
    \includegraphics[width=1\columnwidth]{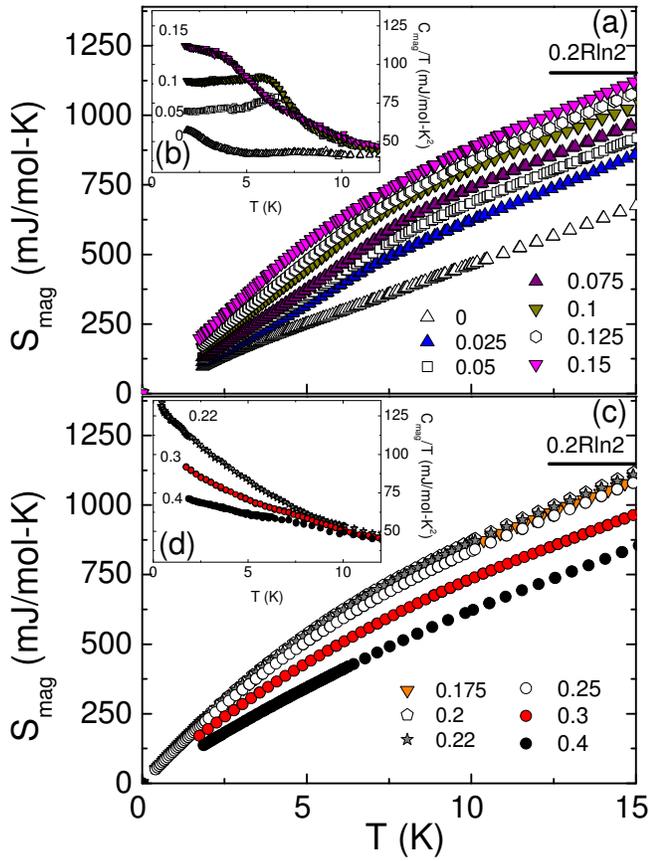}
  \end{center}
  \caption{(a) Magnetic contribution to the entropy $S_{mag}$ versus temperature $T$ for concentrations 0 $\leq$ $x$ $\leq$ 0.15. (b) Magnetic contribution to the specific heat divided by temperature $C_{mag}/T$ versus $T$ for selected concentrations 0 $\leq$ $x$ $\leq$ 0.15. (c) $S_{mag}(T)$ for 0.175 $\leq$ $x$ $\leq$ 0.4. (d) $C_{mag}(T)/T$ for 0.175 $\leq$ $x$ $\leq$ 0.4. }
  \label{fig:fig4}
\end{figure}

The magnetic contribution to the entropy $S_{mag}$ is shown in Figures~\ref{fig:fig4}(a,c), where $S_{mag}$ was obtained by subtracting the lattice contribution ($C_{lat}/T$ $=$ $\beta$$T^2$ for $T_C$ $\leq$ $T$ $\leq$ 15 K with $\beta$ $\approx$ 0.55 mJ/mol-K$^4$) from $C/T$, extrapolating the resulting $C_{mag}/T$ to zero $T$, and integrating between 0 $\leq$ $T$ $\leq$ 15 K. Results for $C_{mag}/T$ for selected samples are shown in Figures~\ref{fig:fig4}(b,d). As summarized in Figure~\ref{fig:fig7}(d), $S_{mag}$(10 K) increases from 0.07Rln2 (R $=$ 8.314 J/mol-K is the ideal gas constant) at $x$ $=$ 0 to 0.15Rln2 at $x_{cr}$ $=$ 0.22, and finally decreases to 0.1Rln2 at $x$ $=$ 0.4, suggesting a build up of entropy near $x_{cr}$. It is noteworthy that peaks in isotherms of entropy versus substituent composition near the QCP were previously observed in the CeRh$_{1-x}$Co$_x$In$_5$ system.\cite{Maple:10,jeffries_2005_1} We also find that $S_{mag}(T_C)$ evolves from 0.03Rln2 for $x$ $=$ 0, through a peak of 0.12Rln2 near $x$ $=$ 0.1, and returns to 0.047Rln2 for $x$ $=$ 0.2, supporting the viewpoint that in the ordered state, Fe substitution initially suppresses the conduction electron itineracy near $x$ $=$ 0.1, after which strong itineracy returns near $x_{cr}$.

\begin{figure}
  \begin{center}
    \includegraphics[width=1\columnwidth]{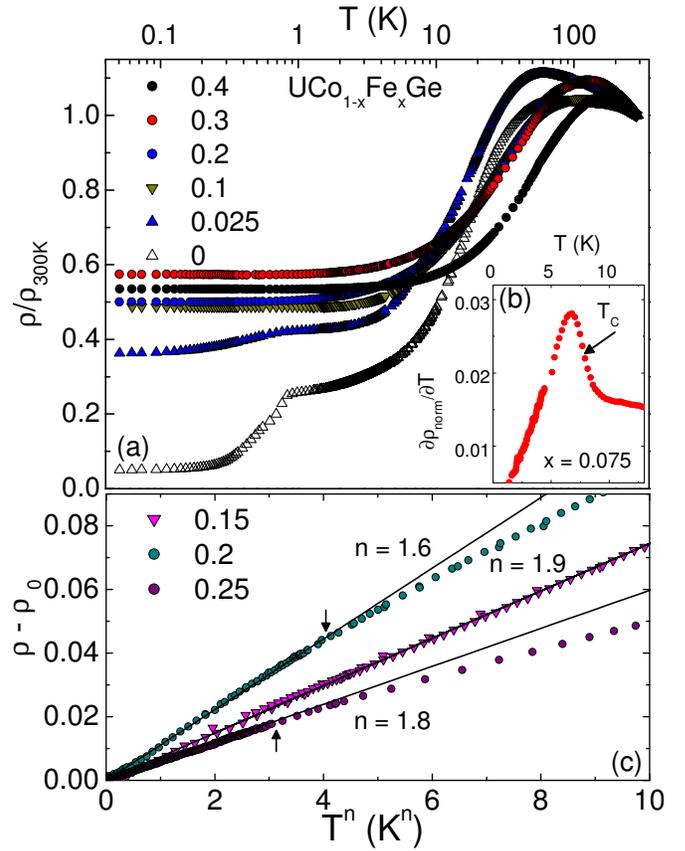}
  \end{center}
  \caption{(a) Electrical resistivity $\rho$ normalized to $\rho_{300K}$ versus temperature $T$ for selected concentrations $x$ on a log scale.  (b) Derivative of electrical resistivity with respect to temperature $\partial$$\rho$/$\partial$$T$ versus $T$ for $x$ $=$ 0.075. (c) $\rho$ minus residual resistivity $\rho _0$ versus $T^n$ for concentrations spanning $x_{cr}$, as described in the text. The solid lines show the power law fits to the data. The black arrows denote the fit range.}
  \label{fig:fig5}
\end{figure}

Representative normalized electrical resistivity $\rho (T)/\rho_{300K}$ data are presented in Figure~\ref{fig:fig5} on a log scale. The residual resistivity $\rho_0$ changes from 70 $\mu$$\Omega$cm for $x$ $=$ 0 to $\approx$ 1 m$\Omega$cm for all other doping concentrations. All samples exhibit metallic behavior with a broad maximum at high temperatures ($T^*$) that is associated with the onset of the coherent ground state. $T^*$ gradually increases with increasing $x$, indicating that the hybridization strength between the $f$- and conduction electrons is enhanced by Fe substitution. As shown in Figure ~\ref{fig:fig7}(e), chemical substitution rapidly introduces disorder, resulting in a RRR $=$ $\rho_{300K}$/$\rho_0$ $\approx$ 2 for the entire doping series. Samples with $0.025 \leq x \leq 0.125$ display peaks in $\partial$$\rho$/$\partial$$T$ at $T_C$, as shown for $x$ $=$ 0.075 in Figure~\ref{fig:fig5}(b), which agree with results from $M(T)$ and $C(T)$. For $T_c$ $\leq$ $T$ $\leq$ $T_C$ and 0 $\leq$ $x$ $\leq$ 0.2, power law behavior of the form $\rho (T) = \rho _0 + AT^n$ is observed [representative curves are shown in Figure~\ref{fig:fig5}(c)].  Power law behavior is also observed for $x$ $=$ 0.2 and 0.22 for $T$ $\leq$ 4 K and 2 K, respectively. The best fits are obtained by plotting $\rho$ versus $T^n$ and adjusting the value of $n$ to maximize the range of linear behavior. The resulting values $n$ are plotted in Figure~\ref{fig:fig7}(e), where NFL-like deviations from $T^2$ behavior start near $x$ $=$ 0.15. Interestingly, it is not possible to extend the power law behavior to the lowest temperature for $x$ $\geq$ 0.3.  Rather, the resistivity weakly increases with decreasing $T$. The origin of this behavior is unclear. Superconductivity appears as nearly complete, and partial resistive transitions for the $x = 0$ and $x = 0.025$ samples, respectively, while no trace of superconductivity is observed at any of the other iron concentrations.

\begin{figure}
  \begin{center}
    \includegraphics[width=\columnwidth]{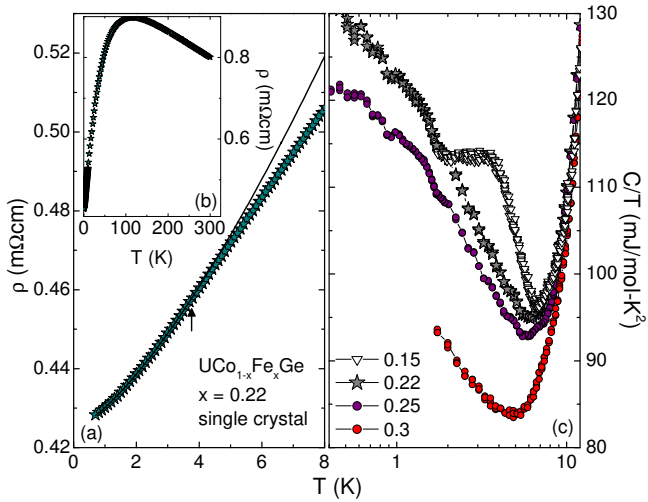}
  \end{center}
  \caption{(a) Electrical resistivity $\rho$ versus temperature $T$ at low $T$ for a single crystal specimen of UCo$_{1-x}$Fe$_x$Ge with $x$ $=$ 0.22. The solid line is a fit to the data of the form $\rho(T)$ $=$ $\rho_0$ $+$ $AT^n$ with $n$ $=$ 1.4, as described in the text. (b) $\rho(T)$ for UCo$_{1-x}$Fe$_x$Ge with $x$ $=$ 0.22 for 0.7 K $\leq$ $T$ $\leq$ 300 K. (c) Specific heat $C$ divided by temperature $T$ versus $T$ for several polycrystalline UCo$_{1-x}$Fe$_x$Ge specimens and the single crystal specimen with $x$ $=$ 0.22. The black arrow denotes the fit range.}
  \label{fig:fig6}
\end{figure}

Having identified evidence for non-Fermi-liquid behavior in the region $x$ $=$ 0.2-0.25, we subsequently synthesized a single crystal specimen with $x$ $=$ 0.22. Results for $\rho(T)$ and $C(T)/T$ are summarized in Figure~\ref{fig:fig6} where we find that the single crystal sample follows the trend that we expect from measurements performed on polycrystalline material. As shown in Figure~\ref{fig:fig6}(a), the electrical resistivity is metallic, has a maximum at $T^*$, and exhibits power law behavior at low $T$ where $n$ $=$ 1.4. Specific heat measurements reveal a nearly logarithmic divergence at low $T$ which is spanned by the low $T$ upturns for $x$ = 0.2 and 0.25. These results strongly suggest the presence of a QCP near $x$ $=$ 0.22 which drives the observed NFL behavior.

\section{Discussion}
\begin{figure}
  \begin{center}
    \includegraphics[width=1\columnwidth]{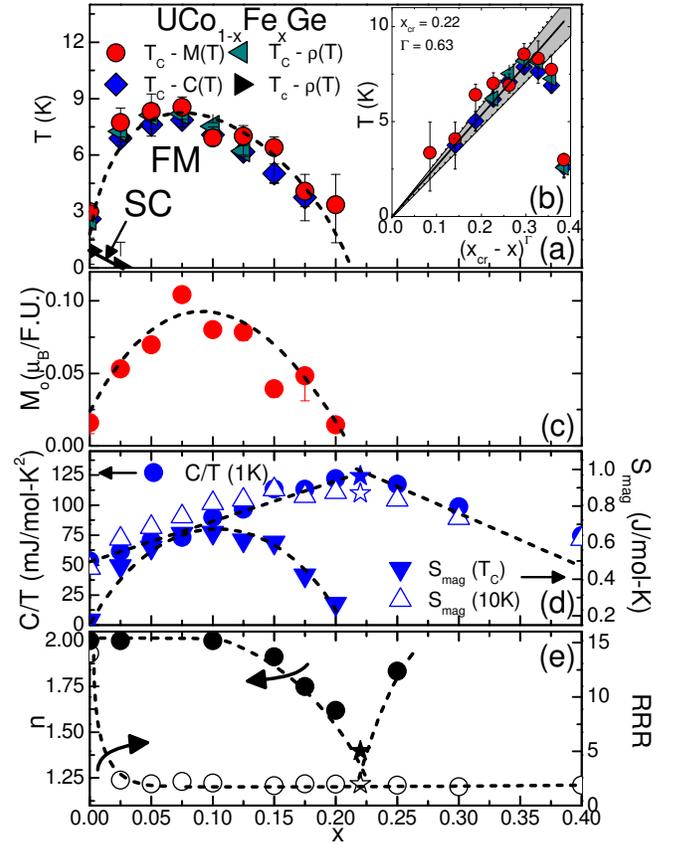}
  \end{center}
  \caption{Iron concentration $x$ dependence of: (a) Curie temperature $T_C$ as determined from magnetization $M$, specific heat $C$, and electrical resistivity $\rho$ data and onset superconducting critical temperature $T_s$ from electrical resistivity; (b) Scaling analysis of the Curie temperature $T_C$, as described in the text. The grey region represents the error in the critical concentration $x_{cr}$ and the exponent $\Gamma$; (c) ordered magnetic moment $M _{o}$; (d) $C/T$ at $T$ $=$ 1 K. For data sets where $C/T$ was measured to 1.8 K, the value at 1 K was estimated by linear extrapolation.; (e) the parameter $n$ obtained from $\rho (T) = \rho _0 + AT^n$ power law fits to the data and the residual resistivity ratio RRR $=$ $\rho_{300K}$/$\rho_0$.}
  \label{fig:fig7}
\end{figure}

Our results reveal an interesting phase diagram for UCo$_{1-x}$Fe$_x$Ge. Initially, ferromagnetism and superconductivity persists over a limited range in $x$. We note, however, that the resistive superconducting transitions are not complete, indicating that the samples investigated in this study do not exhibit bulk superconductivity. With increasing $x$, the ferromagnetic ordering temperature increases and the magnetism becomes more localized, as reflected in several quantities. For 0 $\leq$ $x$ $\leq$ 0.1, electrical resistivity measurements reveal Fermi liquid behavior as evidenced by the $\rho(T)$ $\sim$ $T^2$ behavior. Near $x$ $=$ 0.075, $T_C$ goes through a maximum and subsequently decreases with increasing $x$ until it disappears between $x_{cr}$ $=$ 0.2 - 0.25. In the vicinity of $x_{cr}$, measurements of $M(T)$, $C(T)/T$, and $\rho(T)$ provide evidence for a QCP: (1) $C/T$ obeys a nearly logarithmic divergence in $T$, (2) a maximum in the electronic coefficient of the specific heat is seen around $x_{cr}$, (3) $S_{mag}$ goes through a maximum near $x_{cr}$, and (4) power law fits to $\rho(T)$ reveal that $n(x)$ evolves from Fermi liquid behavior at low $x$ through a ``V-shaped" region with a minimum between $x_{cr}$ $=$ 0.2 - 0.25 and recovers towards $n$ $=$ 2 for larger $x$. We also consider that for a clean itinerant ferromagnetic QCP scenario, it is expected that the ordering temperature should vary as $T_C$ $\sim$ ($x_c$ $-$ $x$)$^{\Gamma}$ where $\Gamma$ $=$ 3/4.\cite{hertz_1976,millis_1993}  As shown in Figure~\ref{fig:fig7}(b), the suppression of $T_C$ is consistent with this prediction, giving $x_{cr}$ $=$ 0.22 $\pm$ 0.01 and $\Gamma$ $\approx$ 0.63 $\pm$ 0.07. Altogether, these results agree with the theoretical prediction for a three dimensional itinerant electron ferromagnetic QPT, except that $n_{cr}$ and $\Gamma$ are slightly reduced from the predicted exponents of 5/3 and 3/4, which are taken from the Moriya SCR and Hertz-Millis theories, respectively.\cite{stewart_2001,moriya_1995,hertz_1976,millis_1993} However, disorder undoubtedly plays a role in determining the ground state behavior in this system and likely perturbs the predicted behaviors, as is also seen for URh$_{1-x}$Ru$_x$Ge and CeSi$_{1.81}$.~\cite{huy_2007_1,drotziger_2006_1,rosch_1999_1} We additionally speculate that the slight saturation in $C/T$ for $x$ $=$ 0.22 occurs either due to disorder or being slightly away from the ideal $x_{cr}$.

Since the question of whether ferromagnetic QPTs are generally first-order as suggested by theory, or second-order as often indicated by experiment, is still a matter of debate, we discuss this issue in more detail. Notably, it is clear that in chemical substitution studies, in particular, disorder may play an important role. While the analysis of our data is in agreement with a scenario involving a clean itinerant ferromagnetic QCP, it is possible to obtain similar NFL behavior due to disorder.\cite{miranda_1997} From this perspective, the QPT at $x_{cr}$ may be a first-order phase transition that is washed out by disorder. We note, however, that similar studies of compounds with comparable RRRs (and thus a comparable amount of disorder) such as UCoGe$_{1-x}$Si$_{x}$ (RRR $\approx$ 4)\cite{ucoge:nijs_1} and URh$_{1-x}$Ru$_{x}$Ge (RRR $\approx$ 2)\cite{huy_2007_1} have usually concluded that the QPT is of second-order. More detailed studies that go beyond the scope of this work that can access the amount of disorder by determining discrepancies between the local structure and average crystal structure, such as EXAFS\cite{booth_2001} and PDF\cite{egami_2003} may ultimately allow this issue to be clarified.

A primary question is why samples in the vicinity of $x = 0.22$ do not display SC, in analogy to the undoped parent compound. One possibility is that if UCoGe exhibits spin triplet SC, which only survives in samples with mean-free-paths that are significantly longer than the coherence length, then disorder should rapidly suppress $T_c$.\cite{Fay:80} If disorder is the main difference between these concentrations, then it also follows that the normal state behavior for $x$ $=$ 0.175 - 0.2 should be similar to that of $x$ $=$ 0. Our results support this viewpoint, particularly for $x$ $=$ 0.175. This relationship is more obvious when we compare to previous pressure and doping studies of UCoGe. Under pressure, $T_C$ decreases and $T_s$ initially increases.\cite{hassinger_2008_1,slooten_2009_1}  Near $1-1.5$ GPa, $T_C$ is suppressed to zero temperature and $T_s$ passes through a weak maximum and subsequently extends beyond the ferromagnetically ordered state.  The magnitude of the feature in the ac magnetic susceptibility at $T_C$, which may be related to the size of the ordered moment, is also gradually suppressed with pressure and extrapolates to zero near $P_c = 1.5$ GPa.  This behavior, together with the possible appearance of NFL behavior in the electrical resistivity near $P_c$,\cite{hassinger_2008_1} suggests the presence of a pressure-induced QPT in UCoGe that may be analogous to what we observe in UCo$_{1-x}$Fe$_x$Ge for $x$ $=$ 0.22. For UCoGe$_{1-x}$Si$_x$, $T_s$ and $T_C$ decrease until they both vanish simultaneously for $x \gtrsim 0.12$ at a QPT that is also accompanied by NFL behavior.\cite{ucoge:nijs_1}  From these observations, we propose that UCoGe, UCoGe$_{1-x}$Si$_x$, and UCo$_{1-x}$Fe$_x$Ge exist on a $T-P-x$ manifold in which the tuning parameters provide access to QPTs which are mainly differentiated by the degree of disorder that is present for each case: i.e., while pressure induces little disorder, Si and Fe substitution result in structural and structural/electronic disorder, respectively, which act to suppress the superconductivity but do not destroy the quantum fluctuations that are responsible for the NFL behavior.

A related challenge is to understand whether the phase transitions near $x$ $=$ 0 and 0.22 are first or second order. It has been proposed that clean metallic ferromagnets with low Curie temperatures in two or three dimensions generically exhibit first order transitions.\cite{belitz_2012_1} This prediction is significant because it differentiates the types of quantum fluctuations that might be found in the vicinity of a FM QPT from those that are found near a continuous QCP. For pure UCoGe, NMR measurements suggest that the ferromagnetic transition is first order.~\cite{hattori_2009_1}  However, bulk measurements such as specific heat and neutron scattering\cite{Stock:11} do not support this viewpoint and, moreover, recent NMR measurements from the same group suggest the existence of critical ferromagnetic fluctuations which drive the superconductivity in UCoGe.\cite{Hattori:12} Therefore, it is not clear whether the ferromagetic phase transition in UCoGe is first or second order. From our measurements, we suggest that the ferromagnetic phase transition near $x$ $=$ 0.22 is second order. Similar behavior is also observed near the critical concentrations in URu$_{2-x}$Re$_x$Si$_2$ and URh$_{1-x}$Ru$_x$Ge, where disorder has been suggested as a mechanism for avoiding a first order transition.~\cite{Butch:09,huy_2007_1,belitz_2012_1} If the nature of the phase transition is different for $x$ $=$ 0 and 0.22, then this may provide an alternative explanation for their differing low $T$ behaviors. Further detailed studies (\textit{e.g.}, inelastic neutron scattering and NMR) of high quality samples near the $x = 0.22$ critical concentration will be useful to reveal differences between the magnetism in undoped UCoGe versus that at $x = 0.22$, which could help to shed light on the precise conditions which lead to SC in UCoGe and NFL behavior in UCo$_{1-x}$Fe$_x$Ge near $x$ $=$ 0.22.

\section{Conclusions}
We have carried out x-ray diffraction, magnetization, specific heat, and electrical resistivity measurements for the UCo$_{1-x}$Fe$_x$Ge series for $0 \leq x \leq 1$, to investigate the $T-x$ phase diagram. Substitution of Fe into UCoGe initially increases the Curie temperature and a rapidly destroys superconductivity.  Near $x=0.22$, the ferromagnetic transition is suppressed to zero temperature at an apparent QCP, where the temperature dependence of the electrical resistivity and specific heat reveal non-Fermi liquid behavior. Having established the $T-x$ phase diagram for this doping series, it will be of great interest to pursue further studies (e.g., neutron scattering, thermal expansion, photoemission, STM) to elucidate the nature of the QPT in this system, as well as the role of disorder.

\section*{Acknowledgements}
Sample synthesis and basic characterization where sponsored by the U.S. Department of Energy (DOE) under Research Grant DE-FG02-04ER46105.  Low temperature dilution refrigerator measurements were supported by the National Science Foundation under Grant DMR-0802478.  MJ gratefully acknowledges financial support by the Alexander von Humboldt foundation. Work at Los Alamos National Laboratory was performed under the auspices of the United States Department of Energy.

\end{document}